# Science Fiction as a Worldwide Phenomenon: A study of International Creation, Consumption and Dissemination

Elysia Celeste Wells

Savannah College of Art and Design
342 Bull St
Savannah, GA 31402 USA
e-mail: ewells21@student.scad.edu

**ABSTRACT**
This paper examines the international nature of science fiction. The focus of this research is to determine whether science fiction is primarily English speaking and Western or global; being created and consumed by people in non-Western, non-English speaking countries? Science fiction's international presence was found in three ways, by network analysis, by examining a online retailer and with a survey. Condor, a program developed by GalaxyAdvisors was used to determine if science fiction is being talked about by non-English speakers. An analysis of the international Amazon.com websites was done to discover if it was being consumed worldwide. A survey was also conducted to see if people had experience with science fiction. All three research methods revealed similar results. Science fiction was found to be international, with science fiction creators originating in different countries and writing in a host of different languages. English and non-English science fiction was being created and consumed all over the world, not just in the English speaking West.

**INTRODUCTION**
The main question addressed in this paper is if science fiction is primarily a English speaking Western phenomenon or are people actively creating science fiction cross culturally and in different languages? This question was asked as part of a larger research project[1], and was asked in order to ascertain if a business tool derived from science fiction has potential global applications. During the course of doing research the author encountered few references to global science fiction. The author observed that most of the science fiction authors, H. G. Wells, Ray Bradbury, Isaac Asimov and so forth that were focused on came from the English speaking western world and science fiction was addressed as a genre of the western world (Rabkin, 1983). *The Cambridge Companion to Science Fiction* almost exclusively focuses on American science fiction (James & Mendlesohn, 2003). Julies Verne is one of the few non-English speaking authors to be named in many of the books reviewed for this research. (Clarke, 1999; James & Mendlesohn, 2003; Kelly et al., 2009, p. 10). Does this mean that science fiction is a genre that is found primarily in the English speaking West[2] or are there science fiction stories being written all around the world that simply have not been addressed by the reviewed literature?

**Literature Review**
In an essay written by James Gunn for *World Literature Today* he states that "American science fiction is the base line against which all the other fantastic literatures in languages other than English must be measured." He goes on to explain "That is because science fiction, as informed readers recognize it today, began in New York City in 1926" (Gunn, 2010) Other sources stated similar things. Science fiction developed during times of great technological growth and change (James & Mendlesohn, 2003). *Science Fiction: A Historical Anthology* states that "[s]cience fiction emerged, quite properly, when science did"(Rabkin, 1983, p. 9). Major American science fiction magazines such as *Astounding Science Fiction and Fact*, *Galaxy Science Fiction* and *Amazing Stories* were founded in the late 1920s and 1930s at the beginning of the nuclear age (IEEE, 2012; Lehrer, 2010). The literature also stated that science fiction and popular science developed together, the founding of the magazine *Popular Science* began when the first science fiction stories were being published (Popular Science, 2002; Rabkin, 1983, p. 221). "The [science fiction] stories printed on pulp had something, besides economics, in common with the newspaper. In a world of conflict made possible by science, the fictions of science became daily fare[3]" (Rabkin, 1983, p. 221).

---

[1] This research was originally done as part of research for a Masters of Fine Arts thesis. This thesis was unpublished at the time when this was written. (Wells, 2013).

[2] For the purpose of this paper the English speaking West is defined as the US, Canada, England, Ireland, Australia and New Zealand.
[3] The term "pulp" refers to inexpensive magazines.

The term science fiction is a modern construction. It was not used until the twentieth century, even for stories that are considered to be science fiction now. Science fiction did not emerge as a named literary genre until the 1930s with the start of "science fiction pulp magazines (James & Mendlesohn, 2003, p. xvi). Wells' stories were "scientific romances" and Jules Verne's were "extraordinary voyages" until they were reclassified and reprinted (Rabkin, 1983, p. 221).

In the essay *Science Fiction Around the World* it was explained that "After World War II, the genre got exported to Western Europe and then, more slowly, to Eastern Europe and the Far East, generally following the progress of industrialization" (Gunn, 2010). The history of science fiction would indicate that the genre is based in the English speaking West.

The subject of science fiction is rarely found as an academic subject outside of literary studies. Despite the history and the popularity of the genre, the scholarship of science fiction did not emerge until the 1950s (James & Mendlesohn, 2003, p. xvii). *Science Fiction Studies* began with night classes but then expanded into scholarly journals and became its own field of study in the 1970s (Kelly et al., 2009, p. 9). The academic field has grown since that time but primarily focuses on literary critique and cultural relevance to modern society. Literature that analyses science fiction itself rather than the subjects and the stories within it are still limited.

## METHODS
Each research method employed was designed to complement the two others. The information gained from the Condor search was used to supplement the Amazon.com data and the international survey to show the global nature of science fiction.

Before starting the research, the expectation was that science fiction was global, but was limited. This expectation was made because of the lack of literature on the subject and because the history of science fiction focused on its origins in the United States and Great Britain. It was discovered that science fiction is indeed global and more wide spread than expected. The Amazon.com data indicated that science fiction is being written by Chinese authors and sold to Chinese audiences; the condor data suggested that science fiction is being talked about in Hindi and in Russian. Most of the survey participants had heard of or consumed science fiction.

**International Survey**
A survey was created to ascertained if people were aware of science fiction. It was created using Google Documents (Writely Team, 2010). The participants were found in a variety of places, most of the participants answered the questions online via posts on personal Facebook walls and on a Facebook page associated with an international student organization. (Facebook.com, 2012). Participants were also found in a Dutch as a second language class, these participants were asked identical questions as the online participants. The survey consisted of mostly open-ended questions. Few personal question were asked because the survey was designed to insure anonymity.

**Amazon.com**
Amazon.com was used to discover if science fiction is being produced and sold worldwide. If multiple Amazon websites are selling science fiction products created by someone who is from the same country as an Amazon site it would infer that science fiction is global and not just a construct of the English speaking Western World.

The online retail site Amazon.com was chosen because of its worldwide presence and because each international site offers some unique goods dependent on the geographical location it is serving (Amazon.com, 2012). The company has websites in nine different countries: the United States of America, Canada, China, France, Germany, Great Britain, Italy, Japan, and Spain.[4] There are potential biases in the data. It should be noted that Amazon.com is based out of the United States, and this might artificially inflate how much American media shows up in the search results,

The top 10 results of each country's Amazon.com website were identified. This was done as a filter in order to find the products that were popular. The top ten results were used to determine if the science fiction being consumed in a particular region was primarily English speaking and Western in origin, or it was based on the native language and culture.

To ensure that results came up in the language of the specific Amazon site, "science fiction" was translated into the language of each of the sites. The "top" results in all categories were searched for using "science fiction" in the appropriate language. Google Translate was used to translate the word science fiction into the different languages (Google Developers, 2012).

---

[4] The international sites were found on "Amazon International: Around The World," published by Amazon inc. (Amazon.com, 2012).

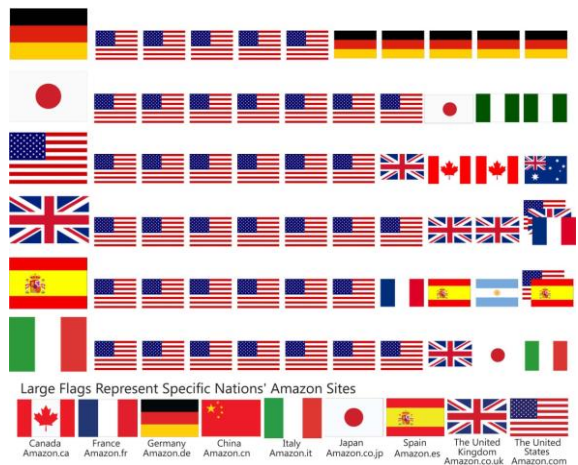

*Figure 1: Are the Writers/Creators of Science Fiction from the same country that their products are sold in?*

**Condor**

In addition to the survey and the Amazon.com results Condor was used (GalaxyAdvisors, 2010). Condor is a program developed by GalaxyAdvisors that is primarily used to locate Galaxy patterns and collaborative networks in databases via an online source (Gloor, 2006, p. 4).

Condor has a "web collector" tool that can locate interconnected websites and networks. This tool enables a broader look at the world wide reach of science fiction because it can look at the World Wide Web, as it existed at the time of the search, to find references of science fiction in multiple locations as it is connected with other terms. The program then creates a visualization of these websites and their connections. For this paper, that visualization shall be the primary source of data.

Collaborative Innovation Networks or COINs are groups of highly self-motivated people who share a collective vision and collaborate with each other. This collaboration leaves a distinctive pattern, as illustrated in figure 2, of interconnected nodes. This "galaxy" appearance occurs because the members communicate with multiple individuals freely instead of one individual being a primary source of communication, as is in the case of hierarchical structures that have the appearance of a star. This is relevant because this signature of cross communication is found in these galaxy patterns. In the case of mapping online websites, galaxy formations appear when they give credit to a different website. For example, if website A cites website B, C, D and F, but websites B, C, D, and F do not reference other websites, or only reference back to website A, you will end up with a star configuration. If website A links to websites B, C, F, H and J, and those websites link not only to each other but to a different group of websites, you get a galaxy formation.

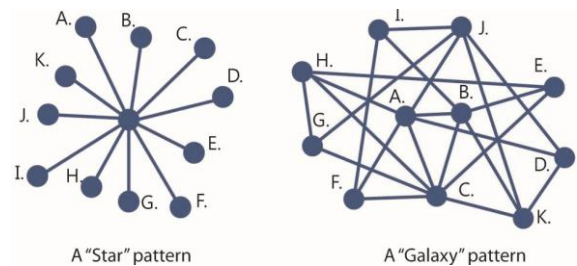

*Figure 2: A Stars Pattern Versus a Galaxy Pattern (Gloor, 2006)*

***What Types of Searches Were Done?***
As mentioned previously, Condor is a computer program that was developed to analyze online networks enabled to find communication networks (GalaxyAdvisors, 2010). Searches using keywords are done to locate communication networks. The network searches for this thesis were done using the Google search engine.

Many searches were conducted but the most relevant ones paired translated terms to look at science fiction in multiple languages. This data filled in the gaps that the Amazon.com. searches and the survey questions left. Multiple languages were used in the search criteria in an attempt to cancel out the English bias. Two words were needed to ensure that the data was more accurate. One term was "science fiction" translated into the target language and the second word used was either "future," again translated into the target language, or the translated version of the word "technology." This was because the word for science fiction is similar in many languages, so the secondary word ensured only results in the targeted language would show up. For example, the French word for science fiction is exactly the same as the English word, and the only difference between the English word and the German word is a dash. The languages used were chosen partially based on the Amazon.com data: the six languages that appeared in the Amazon.com data were used, and additional languages were added if they appeared somewhere in the results. A few languages from areas that were notably absent were also added.

| | | | |
|---|---|---|---|
| English | : science fiction | : technology | : future |
| French | : science fiction | : technologie | : avenir |
| Japanese | : サイエンスフィクション | : テクノロジ | : 将来 |
| Italian | : fantascienza | : tecnologia | : futuro |
| Spanish | : ciencia ficción | : tecnología | : futuro |
| German | : Science-Fiction | : Technologie | : Zukunft |
| Chinese | : 科幻小说 | : 技术 | : 未来 |
| Croatian | : naučna fantastika | : tehnologija | : budućnost |
| Hindi | : साइंस फिक्शन | : प्रौद्योगिकी | : भविष्य |
| Arabic | : القصص الخيالي | : التكنولوجيا | : المستقبل |
| Russian | : научная фантастика | : технология | : будущее |
| Swahili | : Sayansi ya Kubuniwa | : teknolojia | : baadaye |

*Figure 3: Chart of translations*

Using the website Google Translate, science fiction was translated into Arabic, Chinese, Croatian, French, German, Hindi, Italian, Japanese, Russian, Spanish and Swahili (Google Developers, 2012).

**Research Limitations**
The two biggest limitations of the survey were sample size and the language limitations. The survey consisted of 22 individuals. While more would have been optimal, the survey had a large enough sample size to be significant. The second limitation of the survey was that it was primarily in English with a small number of Spanish participants. This could have influenced the results because the majority of the participants had the potential access to English science fiction.

In the Amazon.com research, the biggest limitation was that the retail is currently not truly global. The retailer has not opened up a website in any Africa nations, in India or in Oceania.

## FINDINGS

**Results of the Cultural Survey**
Cultural survey participants were those who grew up in areas that are not English speaking. The vast majority came from areas that would not be considered part of "Western" civilization. This was key in the exploration of the global nature of science fiction, helping to answer the question: Can data from science fiction be used in a global context?

The survey was designed to give insight into how science fiction is perceived by non-English speakers. Are they familiar with the genre? Does their culture have their own version of science fiction?

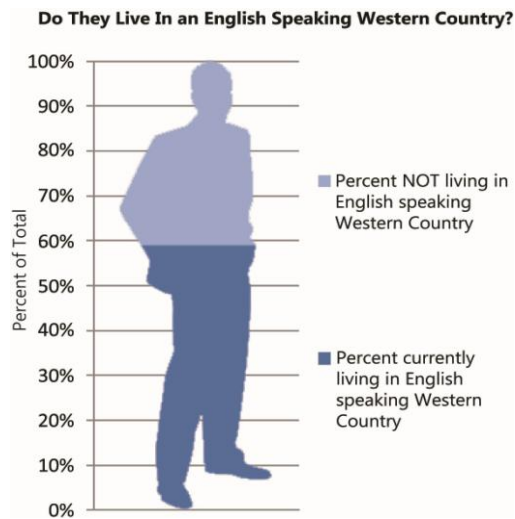

*Figure 4: Do they live where English is spoken?*

Of the 22 people questioned; three came from Europe. The Europeans consisted of one person from Switzerland, one person from Austria and one respondent from "Bosnia and Herzegovina." The largest group came from Asia; One participant came from Vietnam, one from Hong Kong, three from China and one from Indonesia. The second biggest group of people came from Latin American countries; one participant from Honduras, one from Colombia, one from Panama and one participant was from Peru. The rest of the participants came from a variety of places; one participant came from Morocco, one from Iran, two from India, one from Pakistan, one from Nigeria, one from Somalia, one from Uganda, and one participant was from Trinidad.

Approximately 73 percent, or 16 participants were aware of science fiction or somewhat aware of it, and a little more than half reported that they enjoyed science fiction. There was no apparent pattern linking financial situation to exposure to science fiction. Because this was such a small data set, it is still possible that class could affect the likelihood of exposure to science fiction, but people with a variety of income levels were both aware or not aware of the genre, without any obvious correlation.

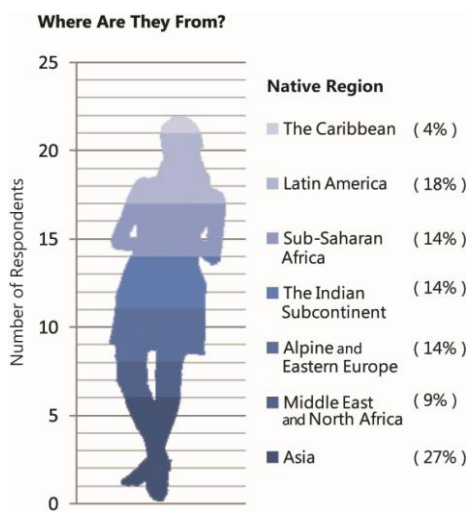

*Figure 5: Where the Respondents are From*

*Does your culture have science fiction?*
One of the first questions in the survey was "Does your culture have stories that show technology that might exist in the future? If yes, can you name some of them?" 13 respondents answered either no, not really, or that they could not think of any. The answer was not consistent among people from the same country: most of the Chinese participants said no, but some said yes and even named a few, including *Little Dragon Boy, - TV Show, Modern Emperor Conflict in China, Future Cops* and *Future X-Cops*. This data implies that exposure to science fiction is varied and while it exists in China, it is not either widely popular or uniformly available.

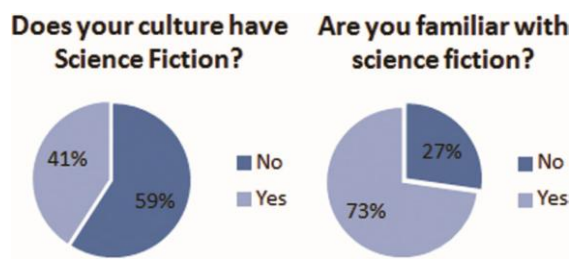

*Figure 6: Are you Familiar with Science Fiction and Does Your Culture Have Science Fiction?*

The participants did repeatedly mention and name "Western" science fiction that they had been exposed to. The question "What science fiction is popular in your country?" was asked. One participant who came from Africa reported, *"*I watched *Back to the Future* growing up in Nigeria, also *Star Trek* on television. I've also watched other movies like *The Minority Report* on DVD." Even though participants could not always name something that is popular in their culture, 16 participants answered yes to the more broad question of, "Are you personally familiar with any science fiction stories?"

**Results of the Amazon.com Analysis**
It was discovered that in all but one market investigated there were products created by a native to that country that was present in the data set. Canada was the only exception. All of the countries, including Canada, had at least one item that was written in the language of that country, and most countries had more than one. This strongly indicates that there is science fiction creation being done in multiple countries and in multiple languages. This was further reinforced by items being created in countries that did not have an Amazon.com website. A box set of DVDs that was being sold on the French Amazon.com site had DVDs that represented three different countries, including Germany and Australia, and the Spanish Amazon..com had a book listed that was written by an Argentinean. Some of the writers whose books were being sold could be listed under two countries; for example, in Spain's top ten list there was a Spanish/American writer who writes in Spanish but teaches at an American University (University of North Texas, 2008).[5]

It should be noted that three of the nine countries did not have science fiction that was created in their country appear in other Amazon.com searches. For example, science fiction from Spain only appeared in the Spanish Amazon.com website search. It is also possible that science fiction is not popular in the regions where there was the data gap; this will be tested later with the Condor research.

It is evident that the dominant country was the United States and the dominant language was English. The flaw of using a website run by an American company is that it is unclear if the large amount of American-created and English-speaking results are because of the website's bias or if there is genuinely a strong desire for these products among people.

**Condor Results**
While condor is an advanced tool, we used it to look for relatively simplistic patterns. Are there numerous nodes, and how connected are the nodes? How thick or thin is the network? The "thickness" or "thinness" of the network was used to determine

---

[5] This information was verified by checking the university's faculty listings.

if science fiction was being discussed about or possibly created in the specific language.

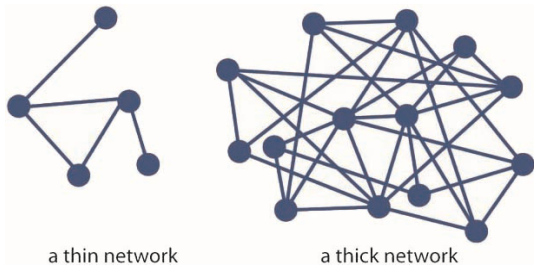

*Figure 7: A Comparison of a Thick vs. Thin Network*

As was expected, some languages had thinner networks than others. The Swahili search results were the thinnest. The Swahili term for science fiction and technology only had one website connected to the term. The Swahili name for science fiction and future was just as thin and also had a single website linked to it.

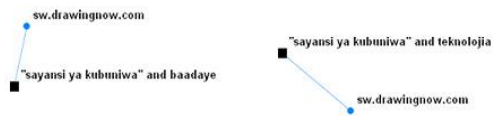

*Figure 8: Swahili language, "Sayansi ya Kubuniwa," "baadaye" and "Teknolojia"* [6]

The website that was found in both searches was the same. This indicated that science fiction, at least online, was not popular in places where Swahili is spoken.

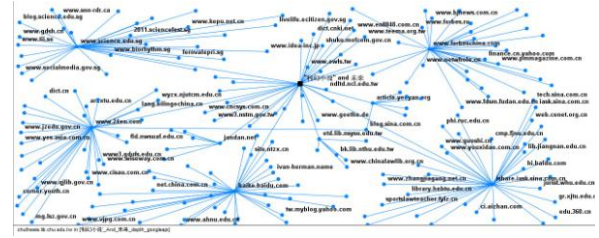

*Figure 9: (left) Science fiction translated into Chinese or "科幻小说" and Future translated into Chinese or "未来."*

Most of the languages have at least one of the two terms, future or technology, which created a similar pattern: primarily sets of stars with a few galaxies. Most of the galaxies were not found in languages spoke in Western European and the Americas such as French, German English, and Spanish. The thickest data sets were found in Croatian and Hindi. This is an indication that science fiction could be global phenomenon.

## CONCLUSION

Overall, this data suggests that science fiction is talked about in different languages and in different places. The survey demonstrated people were aware of science fiction at a global level although unevenly so. It was found that people are referencing science fiction in languages other than English online. The Amazon.com data strongly indicates that science fiction is global by showing non-English speaking Western science fiction is being created and sold to a worldwide audience. While Western English speaking science fiction is also being consumed, it is not the only kind of science fiction being created and sold. The research done for this paper indicates that science fiction is global therefore the data derived from science fiction could be used in a global context.

---

[6] The program "Condor" was used in September 2012. Condor is a program that was created by GalaxyAdvisors for network analysis. The program harvests data from online sources and creates a visual map of that data.